# We should Stop Claiming Generality in our Domain-Specific Language Papers

Daco C. Harkes[a]

a   Delft University of Technology, The Netherlands

**Abstract**    Our community believes that new domain-specific languages should be as general as possible to increase their impact. However, I argue in this essay that we should stop claiming generality for new domain-specific languages. More general domain-specific languages induce more boilerplate code. Moreover, domain-specific languages are co-developed with their applications in practice, and tend to be specific for these applications. Thus, I argue we should stop claiming generality in favor of documenting how domain-specific language based software development is beneficial to the overall software development process. The acceptance criteria for scientific literature should make the same shift: accepting good domain-specific language engineering practice, instead of the next language to rule them all.

## The Art, Science, and Engineering of Programming

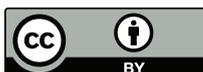

**We should Stop Claiming Generality in our DSL Papers**

## 1 Introduction

Domain-specific languages (DSLs) are thriving in academia and in industry. Our scientific community publishes on average 1.1 articles per day with "domain-specific language" in the title or abstract. At the time of writing Google Scholar lists 418 such publications in the last 365 days. In industry DSL based software engineering is growing rapidly as well. This is illustrated both by the number of DSL talks by companies at academic and industry IT events, and by the bold slogans these companies use: "low-code platform", "auto generated <something>", "automate production of software", and "deliver products 5-10 faster with <DSL-tool>".

Despite advice to not over-generalize [12], scientific arguments for new DSLs in our community usually include an argument for generality. One of the criteria we evaluate DSL research on is its potential impact; and for a DSL to have a large impact, it should be general enough to cover most problems of a domain. In other words, a criterion for accepting DSL research is that a DSL should be widely used potentially.

My goal in this essay is to argue against this criterion, because DSLs can be very useful without generality. This criterion directs research effort in a direction which is not necessarily fruitful. DSLs can become less useful as they become more general, and DSLs which are useful without being general might not be generalizable at all. If despite this, DSL authors still make an argument for generality, it is to be expected that these arguments are weak. They are usually based on anecdotal evidence. Authors list some examples or case studies they could come up with, and which can be expressed in their DSL. Obvious examples which cannot be expressed in the DSL can go to the future work section, but it is best not to list non-obvious limitations because the DSL generality perceived by the reviewers will diminish.

This often leads to claims of generality which are undone by later work. For example a paper stating "we believe that our approach handles many lexically-scoped languages" [13] is followed-up with a paper stating "We also refine and extend our previous theory to address practical concerns including ambiguity checking and support for a wider range of scope relationships" [19]. Similarly, the "optimal-time" [14] incremental evaluation for attribute grammars turned out to be not generally applicable and was improved with "Asymptotic Performance of Symbol Table Modifications" [9]. These weak and false claims for generality leave an impression that work can solve everything. This is harmful for newcomers and dissemination of knowledge to industry as the technical details of a DSL need to be understood before its applicability to a specific problem can be assessed.

We should stop these weak arguments for generality. If authors specifically want to argue for generality, they should have a good methodology for assessing generality. For all other DSL contributions we should stop evaluating generality. Instead, we should report on the whole software engineering process in which DSLs are developed and used, measuring and detailing how DSLs improve the software engineering process and the resulting software.

18:2

## 2 More General DSLs versus more Specific DSLs

First, let us look at generality from a theoretical point of view. Suppose a program can be expressed in a more general DSL and in a more specific DSL, then the program will be more concise in the more specific DSL. A more specific DSL makes assumptions which programs cannot work around, while a more general DSL does not make these assumptions. This means that the more specific DSL cannot express as many programs. However, it also means that programs which do require these assumptions do need to make them explicit in the more general DSL, increasing boilerplate code.

An example of this trade-off between scrap-your-boilerplate and generality is visible in the domain of name binding (figure 1). A very specific DSL for name binding is NaBL [22]. It enables very concise specifications of a small set of static analysis concepts in which scopes are implicit. Scope Graphs [13, 19] make scopes explicit, which enables a wider range of name resolution concepts, but also makes trivial name resolution more verbose. Scope Graphs leave implicit how names flow through scopes. On the generic side we have attribute grammar (AG) based [4] and Datalog-based [16] name resolution which both directly encode how names flow through scopes by means of key-value maps or relations. This enables almost arbitrary name resolution patterns, but increases boilerplate.

It is not necessarily better to have a more general DSL. It can be worth the engineering effort to create a specific DSL for a reasonably complicated program. (Note that it is also not necessarily better to have a more specific DSL, for a relatively simple program the extra overhead of creating and maintaining a more specific DSL might be not worth it.)

## 3 Co-development of DSLs and Software

This trade-off between scrap-your-boilerplate and generality tends to favor the former in practice. DSLs are not designed to be general in practice because most DSLs are developed in tandem with the programs written in them. This is apparent in scientific literature and in the projects using DSLs in academia and industry alike.

### 3.1 Every DSL Paper has its own Running Example

Recently, I had the privilege to work on a survey for incremental computing DSLs. Every paper has its own running example, which can either not be expressed in previous work, or cannot be properly incrementally executed in previous work. These running examples are carefully chosen to exercise the key features of the DSL at hand. Moreover, when research branches out, the running examples do not work in other branches. For example, classical attribute grammars [14, 23, 24] are extended in aggregate AGs [9], remote AGs [1], reference AGs (RAGs) [17], and various higher-order AGs [18, 20, 15, 2, 3], but these extensions do not support each others running examples while most authors claim it is a general framework for building compilers. (One such example is that aggregate AGs (1986) [9] optimize edits on map data

```
Block(_): scopes variable

VarDecl(t, x, _): defines variable x of type t

VarRef(x): refers to variable x

VarRef(x) : t
where definition of x : t
```

```
[[ Block(body) ^ (outer_scope) ]] :=
 new block_scope,
 block_scope -P-> outer_scope,
 [[ body ^ (block_scope) ]].

[[ VarDecl(t, x, _) ^ (scope) ]] :=
 Variable{x} <- scope,
 Variable{x} : t.

[[ VarRef(x) ^ (scope) : x_type ]] :=
 Variable{x} -> scope,
 Variable{x} |-> x_def,
 x_def : x_type.
```

```
inh Variable Name.lookupVariable(String name);
eq Block.getStmt(int index).lookupVariable(String name){
 for(int i = 0; i < index; i++)
   if(getStmt(i).isVarDecl(name))
     return (VarDecl)getStmt(i);
 return lookupVariable(name);
}

syn boolean Stmt.isVarDecl(String name) = false;
eq VarDecl.isVarDecl(String name) =
 name().equals(name);

syn ClassDecl Expr.type() = unknownType();
eq VarRef.type() = lookupVariable(name()) != null ?
 lookupVariable(name()).type() : unknownType();
```

**Figure 1** Trade-off between specificity and generality in name binding DSLs. All snippets define how variable references get resolved to variable definitions, and how blocks scope these variables. (Top) NaBL: A very specific DSL for name binding [22]. It enables very concise specifications of a small set of static analysis concepts in which scopes are implicit. (Middle) NaBL2: Scope Graphs make scopes explicit [13, 19], which enables a wider range of name resolution concepts, but also makes trivial name resolution more verbose. NaBL2 leaves implicit how names flow through scopes. (Bottom) JastAdd: Attribute grammar based name resolution which directly encode how names flow through scopes by means of inherited and synthesized attributes [4]. This enables almost arbitrary name resolution patterns.

structures, while RAGs (2012) state as future work addressing this exact problem: "Some practical RAGs, such as JastAddJ, make use of large attribute values for collecting local declarations into a map data structure. [...] This use of large values causes the same kind of dependency imprecision that ordinary AGs suffer from" [17].)

It is conceivable that these DSL authors wanted to build a piece of software (a compiler in this case), and that a better DSL was the way to do it. The piece of software they wanted to build ended up as the running example in their publication. The software was the driving factor behind the DSL research, and the DSL was the means to build this software. Thus, DSLs were co-developed with the running examples written in them.

### 3.2 DSL Projects in Practice

Co-development is also what I see in projects using DSLs, both in academia and industry. In our building, a group of scientific programmers builds web applications with WebDSL [6]. These programmers are also the maintainers of WebDSL. They add new features to WebDSL, or fix bugs, when this is required for one of the applications. In the same way the additions to IceDust [7] in IceDust2 [8] were required to support a specific application.

Also in industry co-development is the pattern. The Dutch tax office builds the Agile Law Execution Factory (Alef) DSL, with which they build the tax declaration application.[1] Similarly, AFAS builds its customer relation management (CRM) system with a DSL, and when they cannot express something they extend their DSL[2]. M-Industries builds their client applications with Alan [11], and whenever they cannot build something they add another sub language to their family of Alan languages. Finally, the use of mbeddr to build a Smart Meter also included creating new language extensions [21]. These are all typical examples of this pattern I observe around me.

We do not believe in the waterfall process anymore for general software development. Neither should we for DSL-based software development. DSL-based development is iterative, just as software development is. And as requirements and application are co-developed in the iterative process, so are DSLs and applications in the DSL-based iterative process.

## 4 Catering for Co-development

If DSLs (specific and general) are co-developed with applications, we as a research community should support and research that process. In our field we currently have two research directions already supporting this process.

---

[1] Presented at the LangDevCon in 2018 (http://langdevcon.org/).
[2] Presented at the Dutch Software Engineering Symposium in 2016 (http://www.sen-symposium.nl/history/2016/program/).

Extensible languages are one way to cater for co-development of language and application. Extensible language developers expect that their DSL is not a perfect fit for their DSL users. These extensible languages can be customized for specific applications. As mentioned, mbeddr is such an extensible language, and during the Smart Meter development new extensions were added [21].

Another way to cater for co-development is to reduce the development effort of DSLs in general. Language Workbenches aim to do this [5]. In order to provide a seamless co-development cycle of language and application changes in the language should directly be visible in the application. This is the research of interactive or live language workbenches [10].

## 5 Scientific Rhetoric Shift

But more importantly, we need a shift in the rhetoric of our scientific DSL community. We should stop claiming generality, and as reviewers we should stop judging based on the criterion of generality. If we drop the criterion of generality, we might be able to do research in directions which can be more fruitful. DSL authors can stop spending effort on creating weak arguments of why their DSL is general, and reviewers do not have to spend effort on reading those arguments anymore.

Instead, we should start claiming co-development is beneficial, and as reviewers we should start judging on benefit in the software engineering process. Thus we need to report on this co-development process with case studies. Real case studies of the software engineering process are costly, they take enormous effort. Maybe we should, as a community, learn from the medical community, who publish medical trials (case studies) without drawing conclusions or claiming anything. Data on the DSL-based software engineering process should be valuable in itself. If the data is valuable in itself, there is no need to claim in every DSL paper that this DSL is better than all previous DSLs.

Let us be honest, and improve our methodology and rhetoric in DSL research. To inspire us to do that, I want to draw attention to a great example. The case study by Völter et al. of mbeddr use in the Smart Meter project details precisely what DSL use brought to the Smart Meter project: "We find that the extensions help significantly with managing the complexity of the software" [21]. Moreover, they detail how they assessed this benefit: "We look at the mbeddr extensions used[…] We qualitatively assess their impact on the complexity of the implementation" [21]. Finally, it is explicit about conclusions which cannot be drawn based on that case study. For example, they cannot compare development effort with a non-DSL Smart Meter, as they did not build a non-DSL version. I would like to see more such research papers.

I envision that with such research papers we, as a community, can improve the DSL-based software engineering process. And, that in 10 years from now, DSLs are as wide-spread as libraries and frameworks are now.

Note that if DSL authors do write papers with a heavy emphasis on language design instead of benefit to the overall software engineering process, then reviewers should reward authors for being explicit about the limits to the applicability of their DSLs.

Because it is hard to establish the applicability of DSLs (authors cannot do multiple industry-size case studies to gather applicability evidence for a single language design paper), I propose to be explicit about the applicability uncertainty by stating lower bound as well as the upper bound: "Our DSL can do X (we did a case study), it cannot do Y because of Z, and we honestly do not know whether it can do A, B, and C". Being explicit about the applicability, will help newcomers to the field and industry alike.

## 6 Conclusion

DSLs are co-developed with the applications that are written in them. Co-developed DSLs are usually more useful when they are specific instead of general. Thus, our community should drop the generality criterion when evaluating DSL research contributions. Instead, we should report more on how DSL-based development is beneficial to the overall development process. If we keep evaluating DSLs based on their generality, we will see many more weak and false arguments for specific DSLs claiming generality. This is not good for our research community. I hold that we should embrace co-development of DSLs with their applications. Perhaps you disagree. If you do, I hope to read your counterargument in a published essay soon.

**Acknowledgements** I thank Hendrik van Antwerpen and Sebastian Erdweg for their valuable input to this essay. Also, I thank the anonymous reviewers for their constructive feedback.

**We should Stop Claiming Generality in our DSL Papers**

**About the author**

**Daco C. Harkes** is a PhD candidate at Delft University of Technology, where he works on domain-specific languages. His current work includes the domain-specific language IceDust, an incremental computing language for the domain of information systems. You can contact him at d.c.harkes@tudelft.nl.

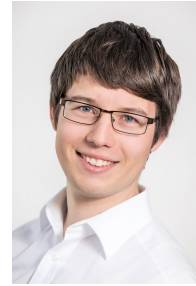